\RequirePackage{fix-cm}
\documentclass[smallextended]{svjour3}       
\smartqed  
\usepackage{graphicx}
\usepackage[scientific-notation=true]{siunitx}
\usepackage{listings}
\usepackage{threeparttable}
\usepackage[bold]{hhtensor}
\usepackage{makecell}
\usepackage{floatrow}
\usepackage[caption=false]{subfig}
\usepackage{threeparttable}
\usepackage[scientific-notation=true]{siunitx}
\usepackage{commath}
\usepackage{chemformula}
\usepackage{booktabs}
\usepackage{cancel}

\begin{document}

\title{Analytical Prediction of Low-Frequency Fluctuations Inside a One-dimensional Shock
}


\author{Saurabh S. Sawant         \and
        Deborah A. Levin  \and         Vassilios Theofilis 
}


\institute{S. S. Sawant \at
              Department of Aerospace, the University of Illinois at Urbana-Champaign, IL, 61801, USA.
              \email{sssawan2@illinois.edu}          
           \and
           D. A. Levin \at
              Department of Aerospace, the University of Illinois at Urbana-Champaign, IL, 61801, USA.
           \and
           V. Theofilis \at
              School of Engineering, University of Liverpool, The Quadrangle, Brownlow Hill, L69 3GH, UK
              Escola Politecnica, Universidade S\~{a}o Paulo, Av. Prof. Mello Moraes 2231, CEP 5508-900, S\~{a}o Paulo-SP, Brasil
}

\date{Received: date / Accepted: date}

\maketitle

\begin{abstract}
Linear instability of high-speed boundary layers is routinely examined assuming quiescent edge conditions, without reference to the internal structure of shocks or to instabilities potentially generated in them.
Our recent work has shown that the kinetically modeled internal nonequilibrium zone of straight shocks away from solid boundaries exhibits low-frequency molecular fluctuations.
The presence of the dominant low frequencies observed using the Direct Simulation Monte Carlo (DSMC) method has been explained as a consequence of the well-known bimodal probability density function (PDF) of the energy of particles inside a shock.
Here, PDFs of particle energies are derived in the upstream and downstream equilibrium regions, as well as inside shocks, and it is shown for the first time that they have the form of the non-central chi-squared (NCCS) distributions.
A linear correlation is proposed to relate the change in the shape of the analytical PDFs as a function of Mach number, within the range $3 \le M \le 10$, with the DSMC-derived average characteristic low-frequency of shocks, as computed in our earlier work.
At a given Mach number $M=7.2$, varying the input translational temperature in the range $89 \le T_{tr,1}/(K) \le 1420$, it is shown that the variation in DSMC-derived low-frequencies is correlated with the change in most-probable-speed inside shocks at the location of maximum bulk velocity gradient.
Using the proposed linear functions, average low-frequencies are estimated within the examined ranges of Mach number and input temperature and a semi-empirical relationship is derived to predict low-frequency oscillations in shocks. Our model can be used to provide realistic physics-based boundary conditions in receptivity and linear stability analysis studies of laminar-turbulent transition in high-speed flows.
\end{abstract}

\section{Introduction}\label{intro}
Supersonic and hypersonic laminar to turbulent flow transition strongly depends on the freestream noise, which includes atmospheric or tunnel-induced turbulence, acoustic, vorticity and entropy fluctuations, and particulates~\cite{schneider2004hypersonic}.
These freestream disturbances interact with a shock and excite the downstream boundary layer through a process known as receptivity~\cite{Morkovin}.
The importance of understanding this process is evident from several experimental works that have shown a strong correlation between the Reynolds number of transition and freestream noise~\cite{potter1968observations,pate1971measurements,schneider2001effects,wagner_sandham_2018}.
Numerical research on the process of receptivity and natural transition has focused primarily on the development of models to account for freestream disturbances~\cite{balakumar2018,hader2018towards} and their interaction with shocks through shock-capturing~\cite{lee1997interaction} or shock-fitting~\cite{zhong1998high} schemes.
However, these works do not account for the internal dynamic structure of a shock-wave which, as our recent work~\cite{sawant2020kinetic} has shown, exhibits {\it low-frequency} molecular fluctuations.
The important role of molecular fluctuations in the laminar to turbulent transition is supported by the works of Fedorov and Tumin~\cite{fedorov2017receptivity} and Luchini~\cite{luchini2010thermodynamic} on the receptivity of boundary layers to equilibrium molecular fluctuations using Landau-Lifshitz's theory of fluctuating hydrodynamics~\cite{landauLifshitz}.
Since freestream noise inevitably passes through a shock before interacting with a boundary-layer, low-frequency fluctuations in the internal translational nonequilibrium zone of a shock may not be ignored in the study of receptivity.
This is true even when the level of freestream noise is minimal, i.e. only containing equilibrium molecular fluctuations, because the leading-edge shock interacts with a boundary-layer formation.
\vspace{\baselineskip}

Some examples of the notable computational fluid dynamic (CFD) literature on modeling the process of natural transition and receptivity include the work of Balakumar and Chou~\cite{balakumar2018}, who introduced the experimentally measured freestream fluctuations by Marineau et al.~\cite{marineau2015investigation} in a Mach 10 flow over a cone at the outer computational boundary of their simulation domain upstream of the detached shock modeled using a shock-capturing scheme.
Hader and Fasel~\cite{hader2018towards} simulated a Mach 6 flow over a flared cone geometry following the experiments in the BAM6QT quiet tunnel at Purdue~\cite{chynoweth2014transition}.
They constructed a simple free-stream noise model and introduced random pressure pertubations in the boundary-layer downstream of the leading-edge shock to account for the effect of freestream disturbances.
Ma and Zhong~\cite{ma2003receptivity1,ma2003receptivity2,ma2005receptivity} studied the receptivity of supersonic boundary layer to four types of freestream disturbances, fast and slow acoustic waves, vorticity waves, and entropy waves. 
The unsteady interaction of the freestream disturbance and the shock was captured using a high-order shock-fitting scheme of Zhong~\cite{zhong1998high}.
\vspace{\baselineskip}

In our work on the low-frequency fluctuations in a one-dimensional (1-D) shock~\cite{sawant2020kinetic}, the shock layer was simulated using the particle-based, high-fidelity DSMC method.
The frequencies of fluctuations in shocks were found to be an order of magnitude lower than those in the freestream for the examined Mach number range, $2 \le M \le 10$.
We showed that this difference results from the well-known bimodal nature of the PDF of gas particles in the shock, as opposed to their Maxwellian distribution in the freestream.
Based on the shape of the DSMC-derived energy distributions at M=7.2, a reduced-order two-bin dynamic model was constructed to account for a large number of collision interactions of particles and fluxes from neighboring zones.
The model correctly predicted the order-of-magnitude differences in frequencies in the shock versus the freestream.
A Strouhal number was also defined based on the bulk velocity upstream of the shock and the shock-thickness based on the maximum density-gradient inside the shock to nondimensionalize the low-frequencies obtained from DSMC.
It remained practically constant when the Mach number was varied by keeping the upstream temperature constant, however, another set of test cases at a constant Mach number of 7.2 and varying input temperature revealed that the Strouhal number decreased with decrease in upstream temperature.
\vspace{\baselineskip}

CFD simulations aimed at understanding receptivity do not model the details of PDFs of particles or the internal structure of a shock wave and therefore, cannot properly account for low-frequency fluctuations.
Yet it is possible to incorporate the effect of characteristic fluctuations of a leading-edge shock by constructing simple models similar to that of Hader and Fasel~\cite{hader2018towards} to understand the process of natural transition in quiet tunnels.
To construct such models, a simple method is needed to estimate the low-frequency of a shock-wave generated at arbitrary input conditions, which may not have been simulated explicity using DSMC.
Also, the Strouhal number may not be the best way to estimate the low-frequency, especially if the input temperature is not the same as that considered in our earlier DSMC-simulations.
In this work, we present a different approach to estimate the characteristic low-frequency of a shock by correlating the shape of the analytically derived PDFs with the DSMC-computed frequencies for a parameter space of $3 \le M \le 10$ and $89 \le T_{tr,1}/(K) \le 1420$.
\vspace{\baselineskip}

The existing literature~\cite{laurendeau2005statistical,vincenti1965introduction,bird:94mgd} on the analytic form of PDFs of energy of particles pertains to systems in global equilibrium with zero bulk velocity, i.e. a heat bath, where the PDFs are of the form of chi-squared distributions.
In this work, we show that the energy PDFs for a hypersonic flow in local equilibrium at the upstream and downstream regions of the shock have the form of the NCCS distributions (see Sect.~\ref{sec:NCCSEq}).
We also derive the bimodal energy PDFs in the nonequilibrium zone of a shock as the linear combination of the NCCS distributions in the upstream and downstream regions with weighting factors obtained from the theory of Mott-Smith~\cite{MottSmith} (see Sect.~\ref{sec:NCCSNonEq}) and provide a python code to easily generate these PDFs (see Appendix~\ref{app:code}).
The variation of the analytically derived PDFs is studied as a function of Mach number and their shape characteristic is linearly correlated with the DSMC-derived low-frequency of the fluctuations inside a shock for the Mach number range, $3 \le M \le 10$ at $T_{tr,1}=710$~K (see Sect.~\ref{sec:correlation}).
At $M$=7.2 and $89 \le T_{tr,1}/(K) \le 1420$, the change in the DSMC-derived low-frequency fluctuation is also linearly correlated with analytically obtained inverse most-probable-speed.
Based on these linear functions, we demonstrate that the average low-frequency of fluctuations can be estimated for any arbitrary input condition within the bounds of the parameter space that is not computed explicitly in our previous work.
For convenience, the estimated average low-frequencies are also provided the entire Mach and temperature ranges.

\section{One-Dimensional Shock Structure and Low-Frequency Fluctuations}~\label{sec:shockAndFluctuations}
\begin{figure}[H]
    \centering
    \includegraphics[width=0.45\textwidth,height=0.40\textwidth]{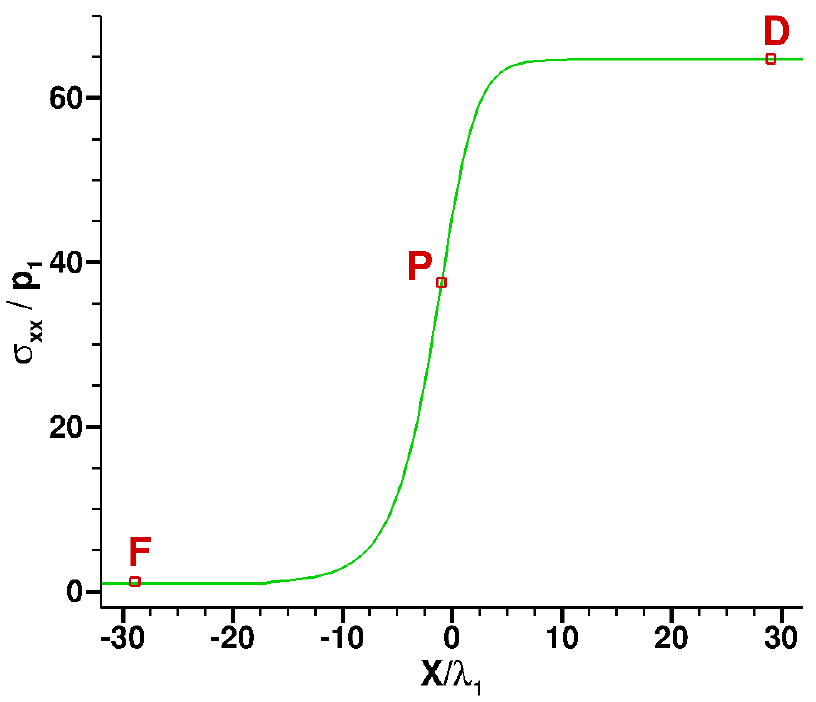}\hfill
    \caption{Normalized $\sigma_{xx}$ in a 1-D, $M$=7.2 shock of argon.}
\label{f:1Dshock}
\end{figure}
The internal structure of a 1-D, steady, Mach 7.2 argon shock simulated using the kinetic DSMC method is shown in figure~\ref{f:1Dshock}, where $X$ is the direction normal to the shock and $\lambda$ is the upstream mean-free-path, and subscripts $1$ and $2$ are used to denote freestream and downstream values.
The upstream number density, $n_1$, and translational temperature, $T_{tr,1}$, are $\num{e22}$~m$^{-3}$ and 710~K, respectively, and the downstream subsonic conditions are imposed by the Rankine-Hugoniot jump conditions.
Figure~\ref{f:1Dshock} shows the $X$-directional normal stress, $\sigma_{xx}$, normalized by the upstream pressure $p_1$, where $\sigma_{xx}$ is non-zero inside the shock, $-16 < X/\lambda_1 < 6$, and zero in the upstream and downstream equilibrium regions, i.e. $X/\lambda_1 \le -16$ and $X/\lambda_1 \ge 6$, respectively.
Numerical probe locations $F$, $P$, and $D$ also marked on the profile of $\sigma_{xx}$ in the freestream ($X/\lambda_1=-29$) region, inside the shock where the gradient of velocity is maximum ($X/\lambda_1=-1$), and in the downstream regions ($X/\lambda_1=29$), respectively.
The complete details of the simulation and profiles of other macroparameters through the shock are given in our recent work~\cite{sawant2020kinetic}.
\vspace{\baselineskip}

To compare molecular fluctuations in the shocks versus that in the freestream, we defined the normalized $X$-directional molecular energy of particles, $\xi_x=(v_x^2-u_x^2)\beta^2$, where $v_x$ and $u_x$ are instantaneous and bulk velocity components of molecules.
$\sigma_{xx}$ is related to $\xi_{x}$ through the first moment of the PDF of $f_{\xi_x}$ defined by mean $\mu_{\xi_x}$ as,
\begin{equation} 
\begin{split}
\mu_{\xi_x}=\int{\xi_x f_{\xi_x}} d\xi_x &= \frac{\sigma_{xx}\beta^2}{\rho}\\
\end{split}
\label{sigmaXXRelation}
\end{equation}
where $f_{\xi_x}$ is the PDF of molecules in the energy space $\xi_x$, $\rho=nm$ is the local mass density, $m=\num{6.63e-26}$~kg is the molecular mass, $\beta=(2RT_{tr})^{-1/2}$ is the the inverse most-probable-speed, $T_{tr}$ is the local average translational temperature, and $R$ is the specific gas constant.
The analytical form of $f_{\xi_x}$ in equilibrium and nonequilibrium regions is described in Sects.~\ref{sec:NCCSEq} and \ref{sec:NCCSNonEq}, respectively.
Reference~\cite{sawant2020kinetic} shows that based on equation~\ref{sigmaXXRelation}, the fluctuations in $\sigma_{xx}$ are related to fluctuations in the number of particles in the energy space, $\xi_{x}$.
\begin{figure}[H]
    \centering
    \sidesubfloat[]{\label{f:FlucAtP}{\includegraphics[width=0.44\textwidth]{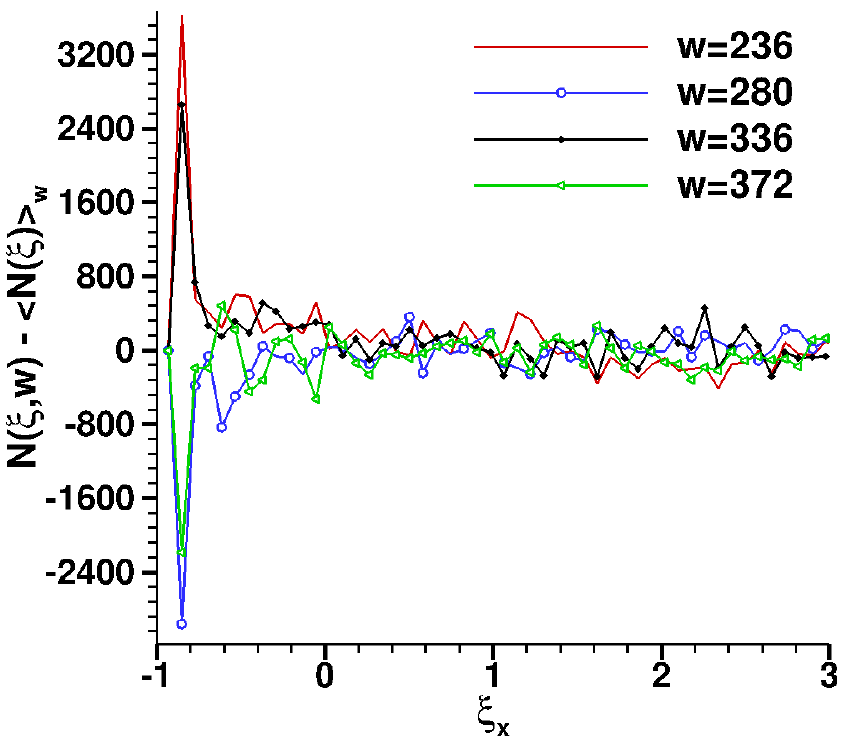}}}\hfill
    \sidesubfloat[]{\label{f:FlucAtF}{\includegraphics[width=0.44\textwidth]{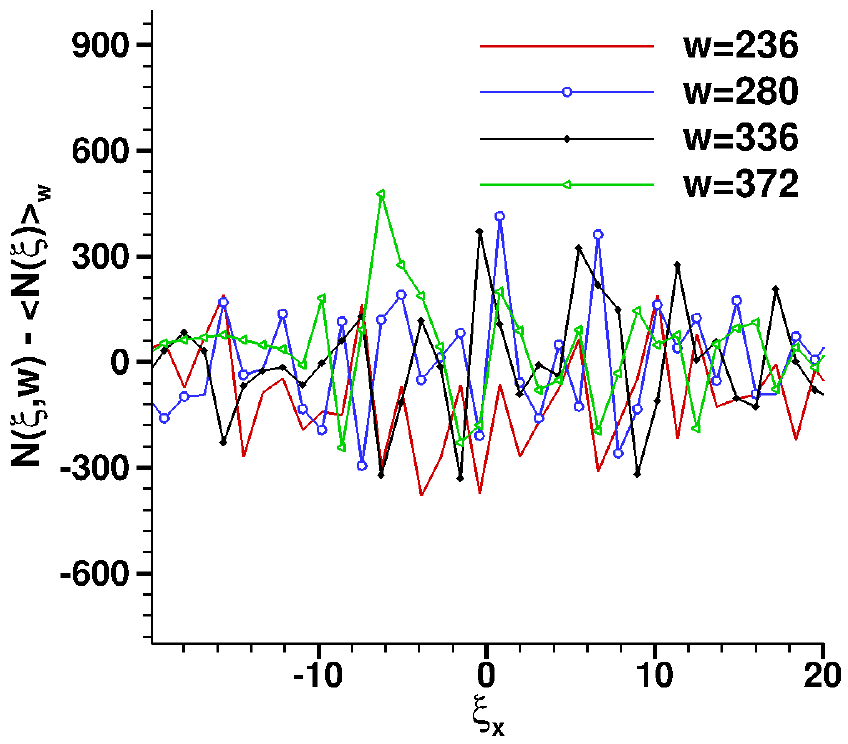}}}\,
    \caption{Fluctuations in the number of particles in PDF of $f_{\xi_x}$ obtained from DSMC as a function of $\xi_x$ at four time-windows (\textit{a}) at probe $P$ inside the shock and (\textit{b}) probe $F$ in the freestream.}
    \label{f:UseOffxix}
\end{figure}

The differences in fluctuations in the shock and the freestream are shown in Figs.~\ref{f:FlucAtP} and~\ref{f:FlucAtF}, respectively.
The fluctuations are defined by the quantity $\left[N(\xi_x,w)-\langle N(\xi_x)\rangle_w\right]$, where $N(\xi_x,w)$ is the total number of DSMC particles within the energy space $\xi_x$ and $\xi_x+\Delta\xi_x$, where $\Delta\xi_x=0.0115$, from time-window $w$ to $w+1$ and $\langle N(\xi_x) \rangle_{w}$ denotes the number of particles in the same energy space averaged over all time-windows, $W=646$.
Each time-window is equal to 100 timesteps of 3~ns and $w$=0 is defined to be the end of transient time of 6~$\mu$s required to establish the shock structure from the initial Rankine-Hugoniot jump conditions.
Figure~\ref{f:FlucAtP} shows at probe $P$ inside the shock, the fluctuations within the energy space $-1$ to $1$ have a time-period of approximately 100 time-windows, corresponding to a 34~kHz frequency.
The low-energy space corresponds to the contribution from the subsonic region.
Note that due to the statistical nature of molecular fluctuations the frequencies are distributed over a broadband of low-frequencies, which is characterised by a weighted-average and $\pm$1 standard deviation. 
At probe $P$, these values were found to be 37.5 and 21.4~kHz, respectively.
The entire spectrum of frequencies obtained using the power spectral density (PSD) of the mean-subtracted, DSMC-derived data of $\sigma_{xx}$ are described in Ref.~\cite{sawant2020kinetic}.
At probe $F$ in the freestream, however, figure~\ref{f:FlucAtF} shows no presence of low-frequencies and the change in the number of particles in freestream is much lower than that in the shock indicating smaller amplitude fluctuations.

\section{Formulation of NCCS PDFs in an Equilibrium Gas}~\label{sec:NCCSEq}
In this section the PDF of $f_{\xi_x}$ is derived for a gas in local equilibrium using the local Maxwell-Boltzmann (MB) PDF~\cite{sharipov2015rarefied} of particle velocities and it is shown that the energy PDF is of the class of an NCCS PDF.
For completeness, PDFs of $f_{\xi_y}$, $f_{\xi_z}$, and $f_{\xi}$ of the $Y$-directional ($\xi_y$), $Z$-directional ($\xi_z$), and total energies of particles ($\xi$) are also derived.
The first moment of $f_{\xi_y}$, $f_{\xi_z}$ and $f_{\xi}$ are related to $Y$ and $Z$-directional normal stresses, $\sigma_{yy}$, $\sigma_{zz}$, and pressure, $p$, as
\begin{equation} 
\begin{split}
\mu_{\xi_y} = \int{\xi_y f_{\xi_y}} d\xi_y &= \frac{\sigma_{yy}\beta^2}{\rho}\\
\mu_{\xi_z} = \int{\xi_z f_{\xi_z}} d\xi_z &= \frac{\sigma_{zz}\beta^2}{\rho} \\
\mu_{\xi  } = \int{\xi f_{\xi  }} d\xi &= \frac{3p\beta^2}{\rho}=\frac{3}{2} \\
\end{split}
\label{momentsOfEnergyPDF}
\end{equation}
Note that $\xi_y=v_y^2\beta^2$, $\xi_z=v_z^2\beta^2$, and $\xi=(v^2-u^2)\beta^2$, where $v_y$, $v_z$ are the instantaneous components in the $Y$ and $Z$-directions, respectively, $u_y$, $u_z$ are the bulk velocity components in the $Y$ and $Z$-directions, respectively, and $v$, $u$ are the instantaneous and bulk speed of molecules, respectively.

The local MB PDF for particles having instantaneous molecular and bulk velocity vectors $\vec{v}$ and $\vec{u}$, respectively, and an equilibrium translational temperature, $T_{tr}$, is written as,
\begin{equation} 
\begin{split}
 f_{\vec{v}} &= \frac{\beta^3}{\pi^{3/2}}\exp{\left(-\beta^2(\vec{v}-\vec{u})^2\right)}\\
\end{split}
\label{MB_vec_v}
\end{equation}
$f_{\vec{v}}$ may change spatially depending on the local values of $\vec{u}$, $T_{tr}$, and applies to particles in equilibrium, i.e. it is valid for $X/\lambda_1 \le -16$ in the freestream and $X/\lambda_1 \ge 6$ in the dowstream.
The bulk velocity and temperature are related to the first and second moments of the distribution as,
\begin{equation} 
\begin{split}
u_i &= \int{v_i f_{\vec{v}}} dv_xdv_ydv_z, \text{\,\,\,\,\,for\,\,}  i~\in \{x, y, z\} \\
T_{tr} &= \frac{m}{3\kappa_b}\int{v^2 f_{\vec{v}}} dv_xdv_ydv_z
\end{split}
\label{MomentsOfMB}
\end{equation}
The PDF for directional velocity components can be obtained by integrating the differential probability $f_{\vec{v}}dv_xdv_ydv_z$ over the remaining two components and written as,
\begin{equation} 
\begin{split}
 f_{v_i} &= \frac{\beta}{\sqrt{\pi}}\exp{\left(-\beta^2(v_i-u_i)^2\right)}; \text{\,\,\,\,\,for\,\,}  i~\in \{x, y, z\}\\
\end{split}
\label{MB_vi}
\end{equation}
Equation~\ref{MB_vi} can be expressed as a PDF of the Gaussian distribution having mean (bulk velocity) $u_i$, and variance $(2\beta^2)^{-1}$ as,
\begin{equation} 
\begin{split}
\mathcal{N}_{v_i}(u_i,(2\beta^2)^{-1}) &= \frac{1}{\sqrt{2\pi(2\beta^2)^{-1}} } \exp{\left(-\frac{\left(v_i-u_i\right)^2}{2(2\beta^2)^{-1}} \right)}\\ 
\end{split}
\nonumber
\end{equation}

This can be further scaled as a PDF of the Gaussian distribution of a scaled variable $\tilde{v}_i$, having unit variance and non-zero scaled mean $\tilde{u}_i$ as,
\begin{equation} 
\begin{split}
\mathcal{N}_{\tilde{v}_i}(\tilde{u}_i, 1) &= \frac{1}{\sqrt{2\pi}} \exp{\left(-\frac{\left(\tilde{v}_i-\tilde{u}_i\right)^2}{2} \right)}\\ 
\end{split}
\label{ScaledGaussian}
\end{equation}
where
$$\tilde{v}_i = \sqrt{2}v_i\beta \text{\,\,\,\,\,and\,\,\,\,\,} \tilde{u}_i = \sqrt{2}u_i\beta $$
Note that the differential probability of particles lying within a velocity space $v_i$ and $v_i + dv_i$ is the same as them being within the scaled velocity space $\tilde{v}_i$ and $\tilde{v}_i + d\tilde{v}_i$, i.e.,
\begin{equation} 
\begin{split}
\mathcal{N}_{v_i}(u_i,(2\beta^2)^{-1}) dv_i &= \mathcal{N}_{\tilde{v}_i}(\tilde{u}_i, 1) d\tilde{v}_i
\end{split}
\nonumber
\end{equation}
From the scaled Gaussian variables $\tilde{v}_x$, $\tilde{v}_y$, and $\tilde{v}_z$, the PDF for the sum of squares of independent combinations of these variables, $\chi$, can be constructed which are known as the noncentral chi-squared distributions~\cite{andras2008properties},
\begin{equation} 
\begin{split}
 f_{\chi} &= \frac{1}{2} \exp{\left(-\frac{\chi + \eta}{2}\right)} \left(\frac{\chi}{\eta}\right)^{\frac{\mathcal{D}}{4}-\frac{1}{2}} I_{\frac{\mathcal{D}}{2}-1}\left(\sqrt{\eta \chi}\right)
\end{split}
\label{GeneralNCCS}
\end{equation}
They have two parameters, the number of degrees of freedom $\mathcal{D}=$1, 2, or 3 depending on the number of independent Gaussian variables used in the sum, $\chi$,
and a noncentrality parameter $\eta$, given by the sum of squares of the respective means of the Gaussian variables ($\tilde{u}_x$, $\tilde{u}_y$, and $\tilde{u}_z$).
Note that $I_{\gamma}(x)$ is the modified Bessel function of the first kind, defined as,
\begin{equation} 
\begin{split}
 I_{\gamma}(x) &= \sum_{m\ge 0} \frac{1}{m! \Gamma(m + \gamma + 1)} \left(\frac{x}{2}\right)^{2m+\gamma}
\end{split}
\nonumber
\end{equation}
The mean, $\mu_{\chi}$, and standard deviation, $\Sigma_{\chi}$, of $f_{\chi}$ is given as,
\begin{equation} 
\begin{split}
\mu_{\chi} &= \mathcal{D} + \eta\\
\Sigma_{\chi} &= \sqrt{2(\mathcal{D} + 2\eta)}
\end{split}
\label{MeanAndStdGeneralNCCS}
\end{equation}
\vspace{\baselineskip}

To derive the PDF of $f_{\xi_x}$, we choose $\chi=\tilde{v}_x^2$.
Since only one independent Gaussian variable is required, we have $\mathcal{D}=1$.
The noncentrality parameter for the given $\chi$ is $\eta=\tilde{u}_x^2$.
By substituting $\chi$, $\mathcal{D}$, and $\eta$ into equation~\ref{GeneralNCCS}, we obtain,
\begin{equation} 
\begin{split}
 f_{\tilde{v}_x^2} &= \frac{1}{2} \exp{\left(-\frac{\tilde{v}_x^2 + \tilde{u}_x^2}{2}\right)} \left(\frac{\tilde{u}_x}{\tilde{v}_x}\right)^{\frac{1}{2}} I_{-\frac{1}{2}}\left(\tilde{u}_x\tilde{v}_x\right)\\
\end{split}
\label{PDF_tilde_v_x^2}
\end{equation}
and using equation~\ref{MeanAndStdGeneralNCCS}, the mean and standard deviation,
\begin{equation} 
\begin{split}
\mu_{\tilde{v}_x^2} &= 1 + \tilde{u}_x^2\\
\Sigma_{\tilde{v}_x^2} &= \sqrt{2 (1 + 2 \tilde{u}_x^2)}
\end{split}
\label{MeanAndStd_tilde_v_x^2}
\end{equation}
The distribution function in equation~\ref{PDF_tilde_v_x^2} can be scaled by a factor of 0.5 to obtain a distribution $f_{\left(\tilde{v}_x^2/2\right)}$ by noting that,
\begin{equation} 
\begin{split}
f_{\tilde{v}_x^2} d\tilde{v}_x^2 = f_{\left(\tilde{v}_x^2/2\right)} d\left(\frac{\tilde{v}_x^2}{2}\right)
\end{split}
\nonumber
\end{equation}
which means that the probability of particles having normalized energies within energy space $\tilde{v}_x^2$ and $\tilde{v}_x^2 + d\tilde{v}_x^2$ remains unchanged.
Therefore, we obtain,
\begin{equation} 
\begin{split}
 f_{\left(\tilde{v}_x^2/2\right)} = 2 f_{\tilde{v}_x^2} &= \exp{\left(-\frac{\tilde{v}_x^2 + \tilde{u}_x^2}{2}\right)} \left(\frac{\tilde{u}_x}{\tilde{v}_x}\right)^{\frac{1}{2}} I_{-\frac{1}{2}}\left(\tilde{u}_x\tilde{v}_x\right)
\end{split}
\end{equation}
This PDF can be shifted by $-\tilde{u}_x^2/2$ to obtain,
\begin{equation} 
\begin{split}
 f_{\left((\tilde{v}_x^2-\tilde{u}_x^2)/2  \right)} &= \exp{\left(-\frac{\tilde{v}_x^2}{2}\right)} \left(\frac{\tilde{u}_x}{\sqrt{\tilde{v}_x^2-\tilde{u}_x^2}}\right)^{\frac{1}{2}} I_{-\frac{1}{2}}\left(\tilde{u}_x\sqrt{\tilde{v}_x^2-\tilde{u}_x^2}\right)
\end{split}
\label{Xi_x_distrib_inVtilde}
\end{equation}
This is the PDF of particles' normalized $X$-directional energy, $\xi_x$, as seen from the following relation.
\begin{equation} 
\begin{split}
\frac{\tilde{v}_x^2-\tilde{u}_x^2}{2} = v_x^2\beta^2 - u_x^2\beta^2 = \xi_x
\end{split}
\nonumber
\end{equation}
Therefore, we can write equation~\ref{Xi_x_distrib_inVtilde} in terms of $\xi_x$ as,
\begin{equation} 
\begin{split}
 f_{\xi_x} &= \exp{\left(-(\xi_x + u_x^2\beta^2)\right)} \left(\frac{u_x\beta}{\sqrt{\xi_x}}\right)^{\frac{1}{2}} I_{-\frac{1}{2}}\left(2u_x\beta \sqrt{\xi_x}\right)
\end{split}
\nonumber
\end{equation}

Finally, the noncentrality parameter can be substituted back where $\eta=\tilde{u}_x^2=2u_x^2\beta^2$ to obtain,
\begin{equation} 
\begin{split}
 f_{\xi_x} &= \exp{\left(-\frac{2\xi_x + \eta}{2}\right)} \left(\frac{\eta}{2\xi_x}\right)^{\frac{1}{4}} I_{-\frac{1}{2}}\left(\sqrt{2\xi_x\eta}\right)
\end{split}
\label{PDF_Xi_x}
\end{equation}
The aforementioned process of scaling the PDF of $f_{\tilde{v}_x^2}$ in equation~\ref{PDF_tilde_v_x^2} scales the mean and standard deviation in equation~\ref{MeanAndStd_tilde_v_x^2} by a factor of half, whereas shifting it further by $-\tilde{u}_x^2/2 = -u_x^2\beta^2$, shifts the mean by the same factor but keeps the standard deviation unchanged.
Therefore, the mean and standard deviation of $f_{\xi_x}$ can be written as,
\begin{equation} 
\begin{split}
\mu_{\xi_x} &= 0.5\\
\Sigma_{\xi_x} &= \frac{\sqrt{2 (1 + 2\eta)}}{2}
\end{split}
\label{MeanAndStd_Xi_x}
\end{equation}

Next, to obtain the PDF of $f_{\xi_y}$, we follow the same procedure, where $\chi=\tilde{v}_y^2$ and $\mathcal{D}=1$, however, note that $\eta=u_y^2=0$.
For $\eta=0$, the noncentral chi-squared distribution in equation~\ref{GeneralNCCS} reduces to a chi-squared distribution and for $\mathcal{D}=1$, it has the form,
\begin{equation} 
\begin{split}
 f_{\chi} &= \sqrt{\frac{\chi}{2\pi}} \exp{\left(-\frac{\chi}{2}\right)}
\end{split}
\label{chisquared}
\end{equation}
as can be verified by first expanding the series of modified Bessel function $I_{-\frac{1}{2}}$ in equation~\ref{GeneralNCCS}, simplifying, and then substituting $\eta=0$.
Additionally, by using the same equation~\ref{MeanAndStdGeneralNCCS}, the constant values of mean and standard deviation of one and 1.414 can be obtained.
By substituting $\chi=\tilde{v}_y^2$ in equation~\ref{chisquared}, we get,
\begin{equation} 
\begin{split}
 f_{\tilde{v}_y^2} &= \frac{\tilde{v}_y}{\sqrt{2\pi}} \exp{\left(-\frac{\tilde{v}_y^2}{2}\right)}
\end{split}
\label{PDF_tilde_v_y^2}
\end{equation}
To obtain the distribution for $\xi_y=v_y^2\beta^2=\tilde{v}_y^2/2$, the distribution $f_{\tilde{v}_y^2}$ can be scaled by a factor of half as,
\begin{equation} 
\begin{split}
 f_{\xi_y} = f_{\left(\tilde{v}_y^2/2\right)} &= \frac{\tilde{v}_y\sqrt{2}}{\sqrt{\pi}} \exp{\left(-\frac{\tilde{v}_y^2}{2}\right)}
\end{split}
\nonumber
\end{equation}
or in terms of $\xi_y$ as,
\begin{equation} 
\begin{split}
 f_{\xi_y} = f_{\left(\tilde{v}_y^2/2\right)} &= 2\sqrt{\frac{\xi_y}{\pi}} \exp{\left(-\xi_y\right)}
\end{split}
\label{PDF_Xi_y}
\end{equation}
The effect of scaling the PDF by a factor of half reduces the mean and standard deviation to 0.5 and $\sqrt{2}$, respectively.
Note that the PDF of $f_{\xi_z}$ has the same form as the PDF of $f_{\xi_y}$ because of axial symmetry in the 1-D case.
\vspace{\baselineskip}

\begin{figure}[H]
    \centering
    \sidesubfloat[]{\label{f:fexy_F}{\includegraphics[width=0.44\textwidth]{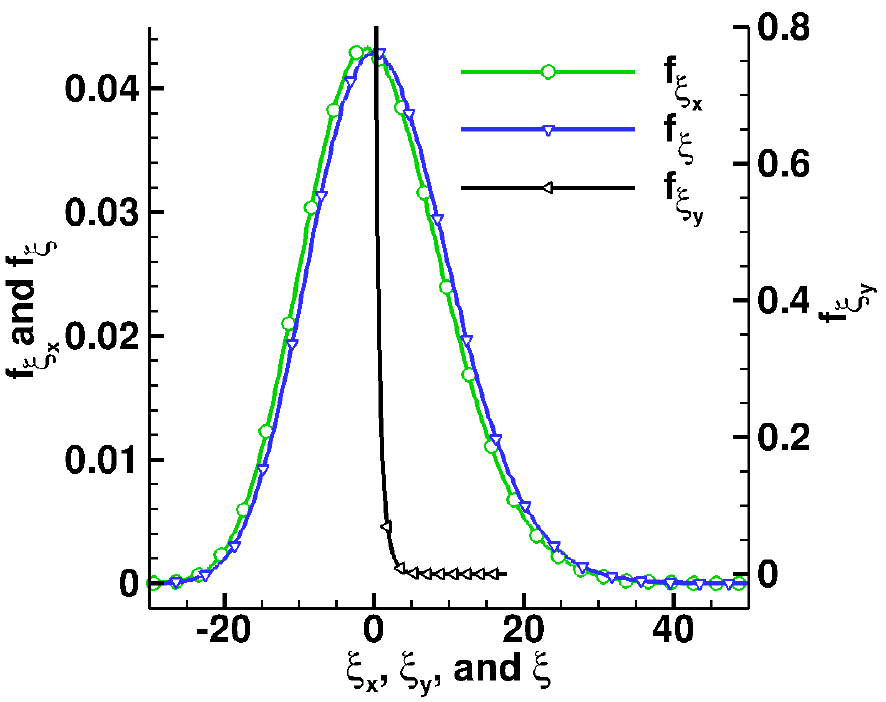}}}\hfill
    \sidesubfloat[]{\label{f:fexy_D}{\includegraphics[width=0.44\textwidth]{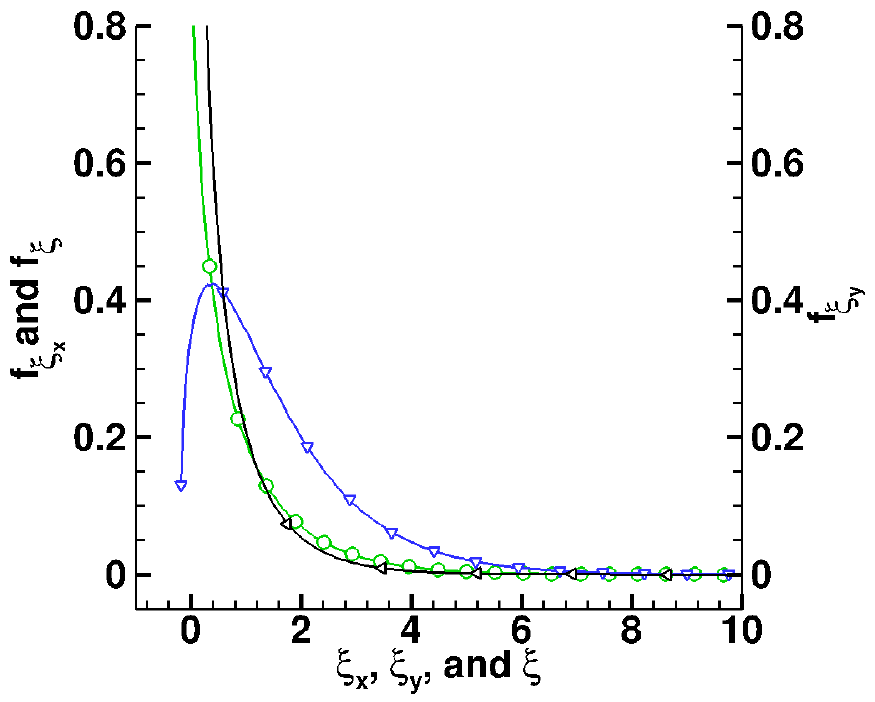}}}\,
    \caption{The theoretically-derived PDFs of $f_{\xi_x}$, $f_{\xi_y}$, and $f_{\xi}$ at (\textit{a}) probe $F$ and (\textit{b}) probe $D$, respectively. Legends for (\textit{b}) are the same as (\textit{a}). The DSMC-derived PDFs are not shown as they agree well with theoretical PDFs.}
    \label{f:DistribProbeFandD}
\end{figure}

Similarly, to obtain the PDF of $\xi=(v^2-u^2)\beta^2$, we choose $\chi=\tilde{v}_x^2 + \tilde{v}_y^2 + \tilde{v}_z^2=\tilde{v}^2$, for which $\mathcal{D}=3$ and $\eta=\tilde{u}_x^2 + \tilde{u}_y^2 + \tilde{u}_z^2=\tilde{u}^2$.
Note that for the 1-D case, $\tilde{u}_y=\tilde{u}_z=0$, which results in $\eta=\tilde{u}_x^2=\tilde{u}^2$.
By substituting $\chi$, $\mathcal{D}$, and $\eta$ in equation~\ref{GeneralNCCS}, we obtain,
\begin{equation} 
\begin{split}
 f_{\tilde{v}^2} &= \frac{1}{2} \exp{\left(-\frac{\tilde{v}^2 + \tilde{u}^2}{2}\right)} \left(\frac{\tilde{v}}{\tilde{u}}\right)^{\frac{1}{2}} I_{\frac{1}{2}}\left(\tilde{u}\tilde{v}\right)
\end{split}
\label{PDF_tilde_v^2}
\end{equation}
By following an approach similar to that for $f_{\xi_x}$, the PDF of $f_{\tilde{v}^2}$ can be first scaled by a factor of half to obtain the PDF of $\left(\tilde{v}^2/2\right)$ and then shifted by $-\tilde{u}^2/2$ to obtain the PDF of $\left((\tilde{v}^2-\tilde{u}^2)/2  \right)=\xi$, which can be written in terms of $\xi$ and $\eta=2u^2\beta^2$ as,
\begin{equation} 
\begin{split}
 f_{\xi} &= \exp{\left(-\frac{2\xi + \eta}{2}\right)} \left(\frac{2\xi}{\eta}\right)^{\frac{1}{4}} I_{\frac{1}{2}}\left(\sqrt{2\xi\eta}\right)
\end{split}
\label{PDF_Xi}
\end{equation}
with mean and standard deviation of $f_{\xi}$ given as,
\begin{equation} 
\begin{split}
\mu_{\xi} &= 1.5\\
\Sigma_{\xi} &= \frac{\sqrt{2 (3 + 2 \eta)}}{2}
\end{split}
\label{MeanAndStd_Xi}
\end{equation}
This completes our derivation of the PDFs of $f_{\xi_x}$, $f_{\xi_y}$, $f_{\xi_z}$, and $f_{\xi}$, which are valid in the equilibrium regions of freestream and downstream.
The mean and standard deviation of these PDFs at probes $F$ in the freestream and $D$ in the downstream are listed in Table~\ref{tab:MeanOfDistrib}.
The theoretical estimates of mean and standard deviation (not shown) agree with values obtained from DSMC-derived PDFs within 1\%, as expected from the good agreement in PDFs shown in Fig.~\ref{f:DistribProbeFandD}.
\begin{table}[H]
\hspace{-2em}
\begin{threeparttable}
\caption{The mean and standard deviation of $f_{\xi}$, $f_{\xi_x}$, and $f_{\xi_y}$ using DSMC.}
\label{tab:MeanOfDistrib}       
\begin{tabular}{|c|c|c|c|c|c|c|c|c|c|}
\hline\noalign{\smallskip}
\textbf{Equation No.}                        &\textbf{Parameters}                          & \textbf{Probes F} & \textbf{Probes D}    \\\hline
        \ref{MeanAndStd_Xi}                  &$\mu_{\xi}  , \Sigma_{\xi}  $     & 1.50, 9.40  & 1.50, 1.36  \\
        \ref{MeanAndStd_Xi_x}                &$\mu_{\xi_x}, \Sigma_{\xi_x}$     & 0.50, 9.35  & 0.50, 0.92  \\
        \ref{MeanAndStd_Xi_x} with $\eta=0$  &$\mu_{\xi_y}, \Sigma_{\xi_y}$     & 0.50, 0.71 & 0.50, 0.72      \\
\noalign{\smallskip}\hline\noalign{\smallskip}
\end{tabular}
\end{threeparttable}
\end{table}

Figures~\ref{f:fexy_F} and~\ref{f:fexy_D} show the PDFs of $f_{\xi_x}$, $f_{\xi_y}$, and $f_{\xi}$ obtained from the analytical expressions given by equations~\ref{PDF_Xi_x},~\ref{PDF_Xi_y}, and~\ref{PDF_Xi} at probes $F$ in the freestream and $D$ in the downstream, respectively.
Since both of these regions are in equilibrium, the theoretically derived distributions agree very well with the numerically obtained distributions from DSMC.
The differences in the respective PDFs at two probe locations are due to the differences in their macroscopic flow parameters.
In the freestream, the PDFs of $f_{\xi_x}$ and $f_{\xi}$ are nearly symmetric about zero, as seen from Fig.~\ref{f:fexy_F}, whereas in the downstream they are almost completely to the right of zero, as seen from Fig.~\ref{f:fexy_D}.
This difference is due to the large noncentrality parameter of $\eta_1=2u_{x,1}^2\beta_1^2=$87.25 in the freestream, for which the NCCS PDF tends towards a Gaussian, whereas the value in the downstream is only $\eta_2=2u_{x,2}^2\beta_2^2=$0.3538.
Parameter $\eta$ can be interpreted as the ratio of the bulk flow energy of the gas to its thermal energy defined using the most-probable-speed, $\beta^{-1}$.
In the downstream, these two energies are expected to be comparable as the kinetic energy of the gas is converted to the thermal energy by the shock.
\vspace{\baselineskip}

Also, the differences between the PDFs of $f_{\xi_x}$ and $f_{\xi}$ are because of the differences in degree, $\mathcal{D}$.
In the freestream, this differences is only noticeable as a small shift between the symmetric distributions, whereas in the downstream the difference in their shapes is quite conspicuous.
Note, however, that at both locations the values of $\mu_{\xi_x}$ and $\mu_{\xi}$ are such that the normalized viscous stress, $\tau_{xx}$, is zero based on the relation,
\begin{equation} 
\begin{split}
\frac{\tau_{xx}\beta^2}{\rho} &= \frac{-(\sigma_{xx}-p)\beta^2}{\rho} = -\left(\mu_{\xi_x} - \frac{1}{3} \mu_{\xi}\right)
\end{split}
\label{ShearStressAndMean}
\end{equation}
which is also consistent with the DSMC-derived profiles of $\tau_{xx}$ (see Ref.~\cite{sawant2020kinetic}).
\vspace{\baselineskip}

Additionally, in the freestream, Fig.~\ref{f:fexy_F} shows that there is a difference between the shape of the PDFs of $f_{\xi_y}$ and $f_{\xi_x}$, although they both have the same degree, $\mathcal{D}=1$.
This difference is because of the difference in $\eta_1$, which is zero for the former and 87.25 for the latter.
In the downstream, these two distributions are very close to each other, as shown in Fig.~\ref{f:fexy_D}, because the value of $\eta_2$ is closer, i.e. 0 and 0.3538 for PDFs of $f_{\xi_y}$ and $f_{\xi_x}$, respectively.
For both equilibrium regions, the mean values $\mu_{\xi_y}$ and $\mu_{\xi}$ are 0.5 and 1.5, respectively, as shown in Table~\ref{tab:MeanOfDistrib}, such that $\tau_{yy}=-(\sigma_{yy}-p)=0$, based on a similar relation as in equation~\ref{ShearStressAndMean}.
Furthermore, notice from Table~\ref{tab:MeanOfDistrib} that the standard deviations $\Sigma_{\xi}$ and $\Sigma_{\xi_x}$ are much larger in the freestream than downstream, because $\eta_1 >> \eta_2$.
Although they are not directly related to any macroscopic flow parameter, they suggest a larger spread of $X$-directional and overall normalized energies of particles in the freestream, where the bulk flow energy is much larger than the thermal energy, as opposed to downstream where the two energies are comparable.
This makes sense because in the downstream, particles collide more and distribute their energy better than the freestream, which causes a reduction in the spread of their energy distribution.

\section{Formulation of Bimodal NCCS PDFs in Shocks}~\label{sec:NCCSNonEq}
To derive the energy PDFs of $f_{\xi_x}$, $f_{\xi_y}$, and $f_{\xi}$ inside the nonequilibrium region of the 1-D shock, $-16 < X/\lambda_1 < 6$, we take the approach of Mott-Smith~\cite{MottSmith}, who approximated the PDF inside the shock of particle velocities as a linear combination of equilibrium PDFs in the upstream and downstream equilibrium regions, i.e.,
\begin{equation} 
\begin{split}
 nf_{\vec{v}} &= \phi_1 f_{\vec{v}, 1} + \phi_2 f_{\vec{v}, 2} 
\end{split}
\label{MB_vi_center}
\end{equation}
where $n$ is the local number density.
PDFs of $f_{\vec{v}, 1}$ and $f_{\vec{v}, 2}$ are obtained by substituting the upstream and downstream macroscopic flow conditions of $(n_1, u_{x,1}, \beta_1)$ and $(n_2, u_{x,2}, \beta_2)$, respectively, in equation~\ref{MB_vec_v}.
The use of normalization condition $\int f_{\vec{v}} d\vec{v}=1$ leads to the local number density, $n=\phi_1 + \phi_2$.
The weights $\phi_1$ and $\phi_2$ are functions of $X/\lambda_1$ and are given as~\cite{bird:94mgd},
\begin{equation} 
\begin{split}
\phi_1 &= \frac{n_1}{1 + \exp{\left(\alpha \left(\frac{X}{\lambda_1}\right)\right)}} \\
\phi_2 &=  \frac{n_2\exp{\left(\alpha \left(\frac{X}{\lambda_1}\right)\right)}}{1+ \exp{\left(\alpha \left(\frac{X}{\lambda_1}\right)\right)}} \\
\end{split}
\label{MB_fractions}
\end{equation}
where, $\alpha$ is a constant given as,
\begin{equation} 
\alpha =  \frac{5}{3\sqrt{\pi}} \frac{\sqrt{2 R T_{tr,1}}}{u_{x,2}} \frac{u_{x,1} - u_{x,2}}{u_{x,1} + u_{x,2}}
\nonumber
\end{equation}
Using the above equations, at probe $P$, the prediction of number density, $n=\phi_1+\phi_2$, is 0.26\% higher than the DSMC-computed value, where $\phi_1$ and $\phi_2$ are $\num{7.214e21}$ and $\num{1.053e22}$, respectively.
By integrating the differential probability $f_{\vec{v}}dv_xdv_ydv_z$ over the remaining two components, the PDF of directional velocities, $v_i$ for $i~\in \{x, y, z\}$ can be obtained inside a shock.
By taking moments of it, the macroscopic flow parameters of directional bulk velocities and directional temperatures can also be obtained as,
\begin{equation} 
\begin{split}
u_i &= \int{v_i f_{v_i}} dv_i, \text{\,\,\,\,\,for\,\,}  i~\in \{x, y, z\} \\
T_{tr,i} &= \frac{m}{\kappa_b}\int{v_i^2 f_{v_i}} dv_i
\end{split}
\label{MomentsOfBimodalMB}
\end{equation}

\begin{figure}[h]
    \centering
    \sidesubfloat[]{\label{f:f_vx_P}{\includegraphics[width=0.44\textwidth]{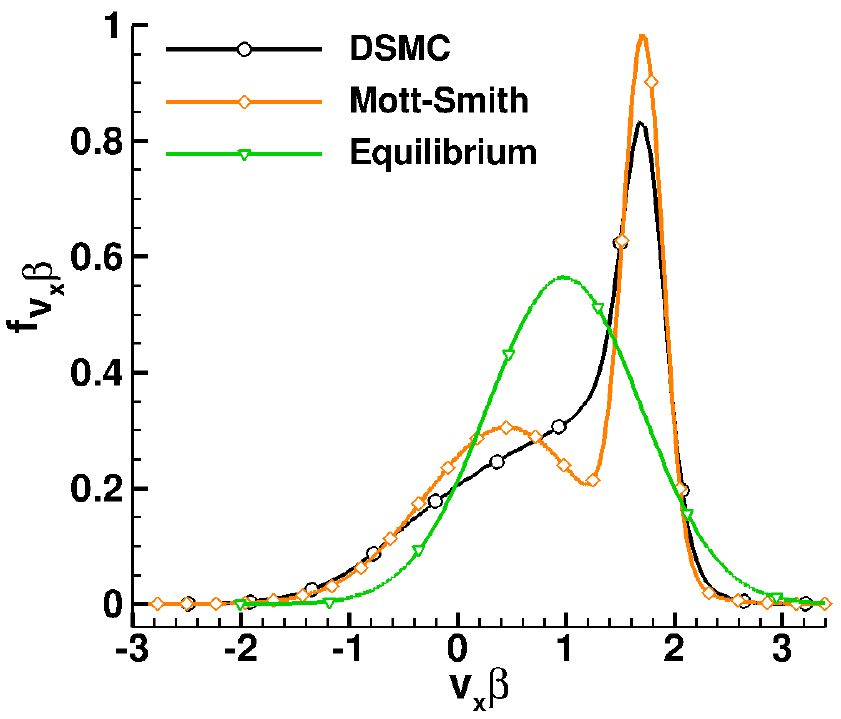}}}\hfill
    \sidesubfloat[]{\label{f:f_vy_P}{\includegraphics[width=0.44\textwidth]{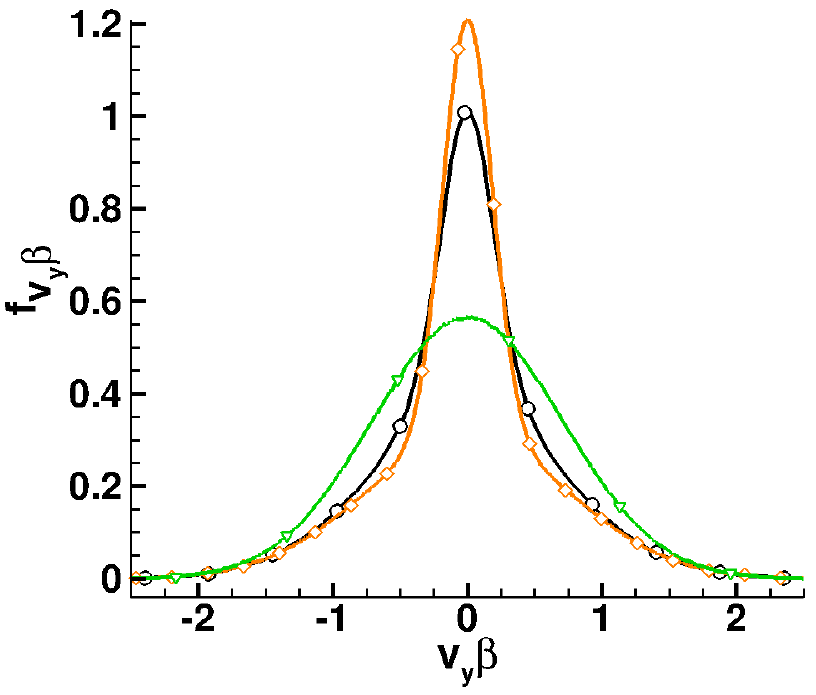}}}\,
    \sidesubfloat[]{\label{f:f_e_P}{\includegraphics[width=0.44\textwidth]{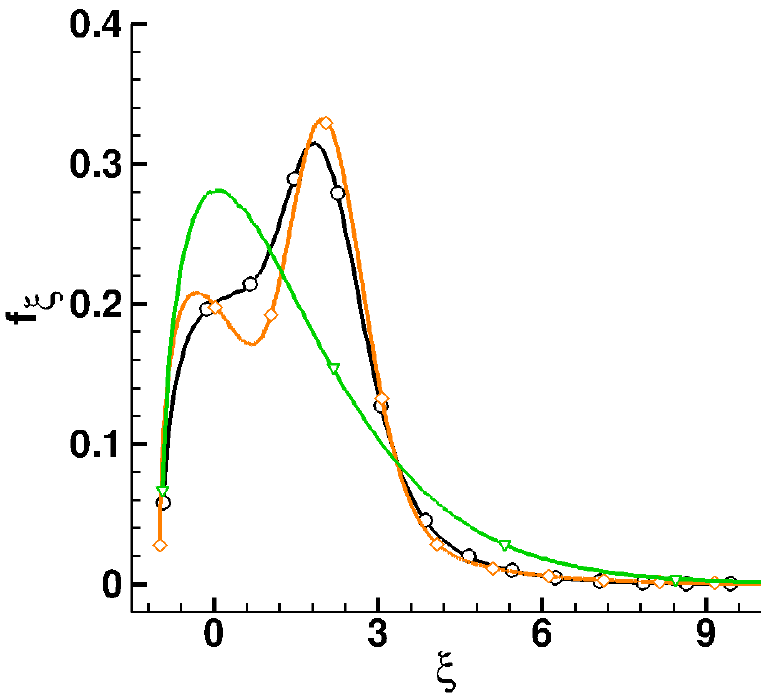}}}\hfill
    \sidesubfloat[]{\label{f:f_ex_P}{\includegraphics[width=0.44\textwidth]{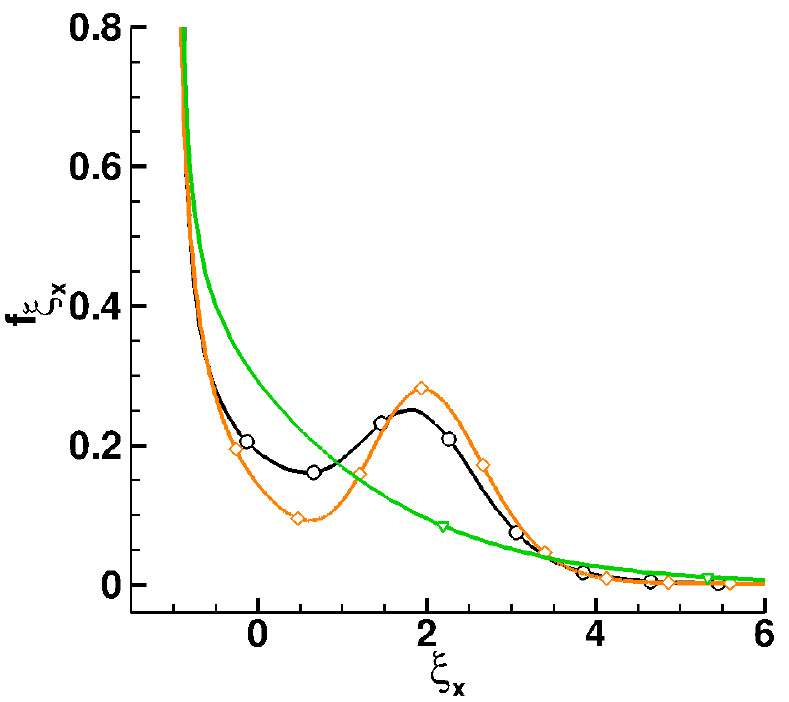}}}\hfill
    \sidesubfloat[]{\label{f:f_ey_P}{\includegraphics[width=0.44\textwidth]{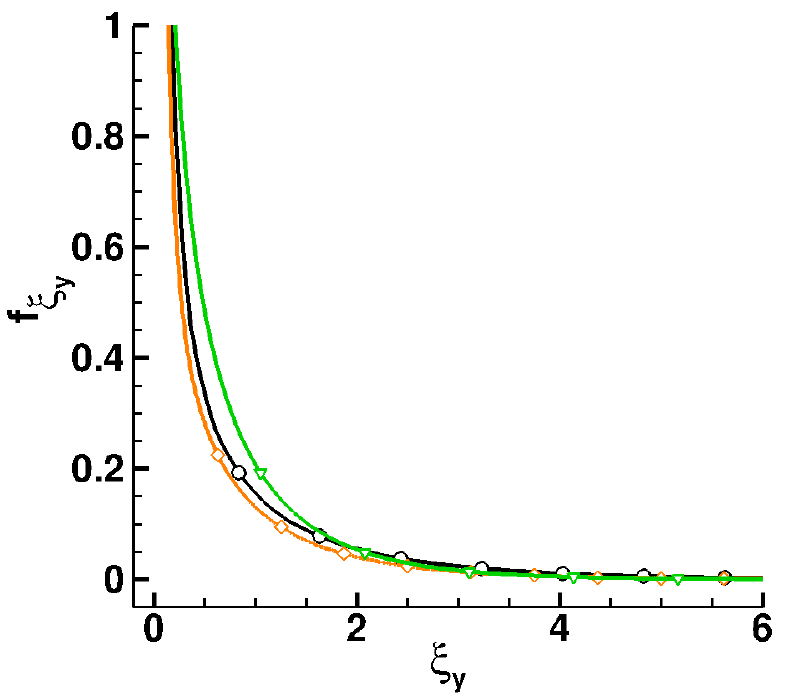}}}\,
    \caption{At probe $P$, the PDFs of normalized velocities (\textit{a}) $v_x\beta$ and (\textit{b}) $v_y\beta$, normalized energies (\textit{c}) $\xi$, (\textit{d}) $\xi_x$, (\textit{e}) $\xi_y=\xi_z$. The DSMC-derived PDFs~\cite{sawant2020kinetic} are compared with the Mott-Smith model and equilibrium theory. The legends shown in (\textit{a}) also apply to (\textit{b})--(\textit{e}). In (\textit{a}) and (\textit{b}), the PDFs of normalized velocities are obtained by scaling velocity distributions as, $f_{v_i\beta}=\beta^{-1}f_{v_i}$. }
    \label{f:DistribProbeP}
\end{figure}

Figures~\ref{f:f_vx_P} and~\ref{f:f_vy_P} show the respective comparison of the PDFs of normalized velocities, $v_x\beta$ and $v_y\beta$, obtained from DSMC with the theoretical prediction of Mott-Smith bimodal model and the equilibrium distribution given by equation~\ref{MB_vi} at probe $P$.
Figure~\ref{f:f_vx_P} shows that the PDF of $v_x\beta$ obtained from DSMC skews towards the high velocity region in contrast to the symmetric unimodal equilibrium distribution.
The Mott-Smith model, gives a qualitatively better approximation in terms of bimodality of the PDF, however, it overestimates the contribution of the upstream high velocity particles and shows a second distinct peak.
This difference leads to a 1.95\% lower prediction of the bulk velocity, $u_x$, in comparison to DSMC at probe $P$, although a smaller difference is expected further away from the center of the shock, where the degree of nonequilibrium is lower.
Figure~\ref{f:f_vy_P} shows a good agreement between Mott-Smith and DSMC except for the region of low $Y$-velocity, $0.25 < v_y\beta < 0.25$, which is overestimated, but nonetheless, the model correctly predicts $u_y=u_z=0$.
The predicted values of directional temperatures, $T_{tr,x}$ and $T_{tr,y}=T_{tr,z}$, are higher and lower than DSMC by 1.35 and 7.2\%, respectively, and the predicted overall temperature, computed as the average of directional temperatures, and the value of $\beta$ are 2.7\% lower and 1.4\% higher than DSMC, respectively.
Since the Mott-Smith model gives a reasonable estimation of macroscopic flow parameters, a similar idea can be used to analytically derive bimodal NCCS PDFs inside the 1-D shock.
\vspace{\baselineskip}

We start with normalizing the weights $\phi_1$ and $\phi_2$ by the total weight $\phi_1+\phi_2$ and use these for constructing a bimodal PDF as,
\begin{equation} 
\begin{split}
 f_{\xi_i} &= \psi_1 f_{\xi_i, 1} + \psi_2 f_{\xi_i, 2} \text{\,\,\,\,\, where, $i~\in \{x, y, z\}$}\\ 
 f_{\xi} &= \psi_1 f_{\xi, 1} + \psi_2 f_{\xi, 2}\\
\end{split}
\label{MottSmith_Xi}
\end{equation}
where
$$ \psi_1 = \frac{\phi_1}{\phi_1+\phi_2} \text{\,\,\,\,\,and\,\,\,\,\,} \psi_2 = \frac{\phi_2}{\phi_1+\phi_2}$$
The PDFs of $f_{\xi_x, 1}$, $f_{\xi_y, 1}$, and $f_{\xi, 1}$ are generated from PDFs of $\tilde{v}_x^2$, $\tilde{v}_y^2$, and $\tilde{v}^2$ given in equation~\ref{PDF_tilde_v_x^2},~\ref{PDF_tilde_v_y^2}, and~\ref{PDF_tilde_v^2}, respectively, by first scaling them by a factor of $s_1=0.5\beta^2/\beta_1^2$ and then shifting by the amounts $-u_x^2\beta^2$, $0$, and $-u^2\beta^2$, respectively.
The local values of $u_x$, $u$, and $\beta$ are obtained from equation~\ref{MomentsOfBimodalMB}.
Similarly, to obtain PDFs of $f_{\xi_x, 2}$, $f_{\xi_y, 2}$, and $f_{\xi, 2}$ the scaling factor of $s_2=0.5\beta^2/\beta_2^2$ is used, while the amount of respective shifting is the same, and the procedure for generating $f_{\xi_z, 1}$ and $f_{\xi_z, 2}$ is the same as that for $f_{\xi_y, 1}$ and $f_{\xi_y, 2}$, respectively, because of axial symmetry.
The final PDFs of $f_{\xi, j}$, $f_{\xi_x, j}$, and $f_{\xi_y, j}$, where $j=1$ and $2$ correspond to upstream and downstream conditions, respectively, are given as,
\begin{equation} 
\begin{split}
 f_{\xi,j} = f_{(v_{x,j}^2-u^2)\beta^2} &= \frac{1}{2s_j}\exp{\left(-2\frac{(v_{x,j}^2-u^2)\beta^2}{s_j} + \frac{\eta_j}{2}\right)}\times\\
             &\left(\frac{4(v_{x,j}^2-u^2)\beta^2 }{s_j\eta_j} \right)^{\frac{1}{4}} I_{\frac{1}{2}}\left(\sqrt{\frac{4\eta_j(v_{x,j}^2-u^2)\beta^2}{s_j}}\right)\\
 f_{\xi_x,j} = f_{(v_{x,j}^2-u_x^2)\beta^2} &= \frac{1}{2s_j}\exp{\left(-2\frac{(v_{x,j}^2-u_x^2)\beta^2}{s_j} + \frac{\eta_j}{2}\right)}\times\\
             &\left(\frac{s_j\eta_j}{4(v_{x,j}^2-u_x^2)\beta^2 }\right)^{\frac{1}{4}} I_{-\frac{1}{2}}\left(\sqrt{\frac{4\eta_j(v_{x,j}^2-u_x^2)\beta^2}{s_j}}\right)\\
 f_{\xi_y,j} = f_{\left(v_y^2\beta^2\right)} &= \frac{1}{s_j}\sqrt{\frac{v_{y,j}\beta^2}{\pi}} \exp{\left(-v_{y,j}\beta^2\right)}\\
\end{split}
\label{PDF_Xi_x_1}
\end{equation}
where, $v_{x,j}$ and $v_{y,j}$ are instantaneous velocities of particles that follow PDFs of $f_{v_x,j}$ and $f_{v_y,j}$ in equation~\ref{MB_vi_center}, for $j=1,2$.
Also, $\eta_j=2u_{x,j}^2\beta_j^2=2u_{j}^2\beta_j^2$, for $j=1,2$.
Note that the mean of these PDFs can be easily calculated as,
\begin{equation} 
\begin{split}
\mu_{\xi} &=   \psi_1 s_1(3+\eta_1) + \psi_2 s_2(3+\eta_2)  - u_x^2\beta^2\\
\mu_{\xi_x} &= \psi_1 s_1(1+\eta_1) + \psi_2 s_2(1+\eta_2)  - u_x^2\beta^2\\
\mu_{\xi_x} &= \psi_1 s_1 + \psi_2 s_2\\
\end{split}
\label{MeanAndStd_Xi_MottSmith}
\end{equation}
however, for higher moments the distributions have to be numerically integrated.
A simple Python code to generate the bimodal distributions of energy using Mott-Smith fractions and its moments is given in Appendix~\ref{app:code}.
\vspace{\baselineskip}

Figures~\ref{f:f_ex_P}, \ref{f:f_ey_P}, and~\ref{f:f_e_P} show the respective comparison of the PDFs of $f_{\xi_x}$, $f_{\xi_y}$, and $f_{\xi}$ obtained from DSMC with the analytically derived bimodal NCCS PDFs using equation~\ref{MottSmith_Xi}, denoted as `Mott-Smith', and the equilibrium NCCS PDFs obtained from equations~\ref{PDF_Xi_x},~\ref{PDF_Xi_y}, and~\ref{PDF_Xi} at local macroscopic flow conditions for probe $P$.
The difference between PDFs obtained from DSMC and the equilibrium theory is expected due to nonequilibrium. 
It is small for PDF of $f_{\xi_y}$, indicating that the energy distribution in transverse directions is not significantly affected. 
The DSMC-derived PDFs of $f_{\xi}$ and $f_{\xi_x}$ exhibit inflection points at $\xi=0.49$ and $\xi_x=1.33$, which indicate their bimodal nature.
\vspace{\baselineskip}

In our previous work~\cite{sawant2020kinetic}, we have used the DSMC-derived PDF of $f_{\xi_x}$ to show that the collisions between particles on two sides of the inflection point are responsible for the low-frequency fluctuations inside a shock.
Figure~\ref{f:DistribProbeP} shows that the Mott-Smith model approximates the bimodality of the PDF reasonably well except for the number of particles in the vicinity of the inflection points.
Despite these differences, Table~\ref{tab:MeanOfDistribBM} shows that the mean and standard deviation of the PDF of $f_{\xi_x}$ obtained from the Mott-Smith model are only 0.6\% lower and 4.4\% higher than the values obtained from the DSMC distribution, respectively, and the mean and standard deviation of the PDF of $f_{\xi_x}$ are 1.3 and 5.8\% higher than DSMC, respectively, whereas for the PDF of $f_{\xi_y}$, these quantities are 7.8 and 9.2\% higher, respectively.
We will ignore these small differences and use the analytical PDF in the next section to correlate its shape with the low-frequency of fluctuations obtained from DSMC.
\begin{table}[H]
\hspace{-2em}
\begin{threeparttable}
\caption{At probe $P$ inside a shock, the mean and standard deviation of $f_{\xi}$, $f_{\xi_x}$, and $f_{\xi_y}$ using DSMC and the Mott-Smith model.}
\label{tab:MeanOfDistribBM}       
\begin{tabular}{|c|c|c|c|c|c|c|c|c|c|}
\hline\noalign{\smallskip}
\textbf{Parameters}              &          \textbf{DSMC}  & \textbf{Mott-Smith}       \\\hline
$\mu_{\xi}  , \Sigma_{\xi}  $    & 1.50, 1.37 & 1.50, 1.43\\
$\mu_{\xi_x}, \Sigma_{\xi_x}$    & 0.75, 1.37 & 0.76, 1.45 \\
$\mu_{\xi_y}, \Sigma_{\xi_y}$    & 0.40, 0.77 & 0.37, 0.71   \\
\noalign{\smallskip}\hline\noalign{\smallskip}
\end{tabular}
\end{threeparttable}
\end{table}

\section{Correlation of the Bimodal NCCS PDF with DSMC-derived Low-Frequency Fluctuations in shocks}~\label{sec:correlation}
This section describes the correlation of the analytical bimodal NCCS PDF of $f_{\xi_x}$ with the DSMC-derived characteristic average low-frequency, $\nu$, in argon shocks, obtained in our previous work~\cite{sawant2020kinetic}.
Using these correlations, an estimate of $\nu$ is made for the Mach number range $3 \le M \le 10$ and  input temperature range $89 \le T_{tr,1} \le 1420$.
\vspace{\baselineskip}

\begin{figure}[H]
    \centering
    \sidesubfloat[]{\label{f:bimodalfxix}{\includegraphics[width=0.44\textwidth]{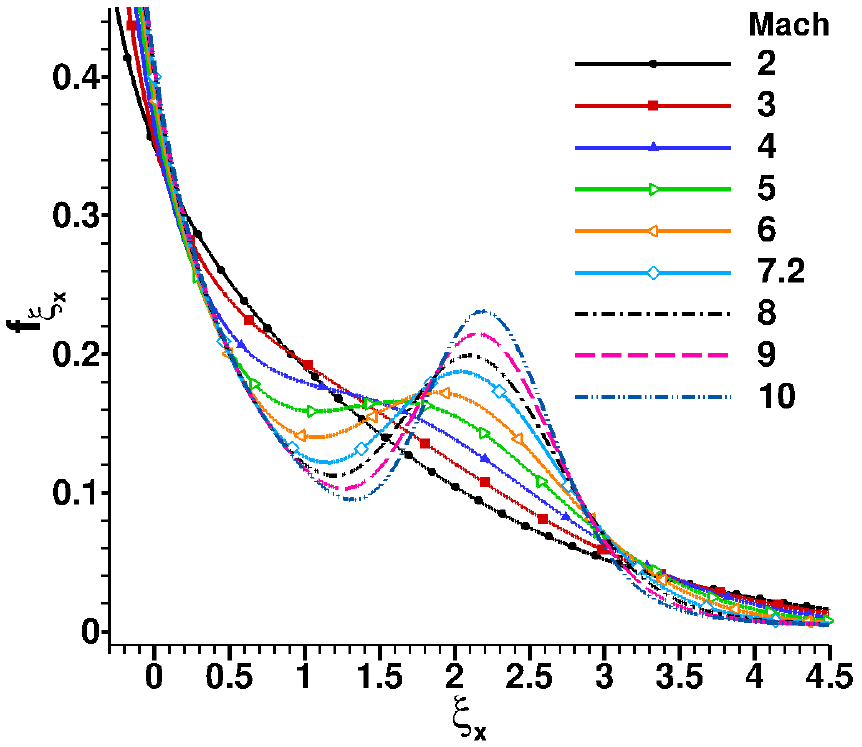}}}\hfill
    \sidesubfloat[]{\label{f:secGrad}{\includegraphics[width=0.44\textwidth]{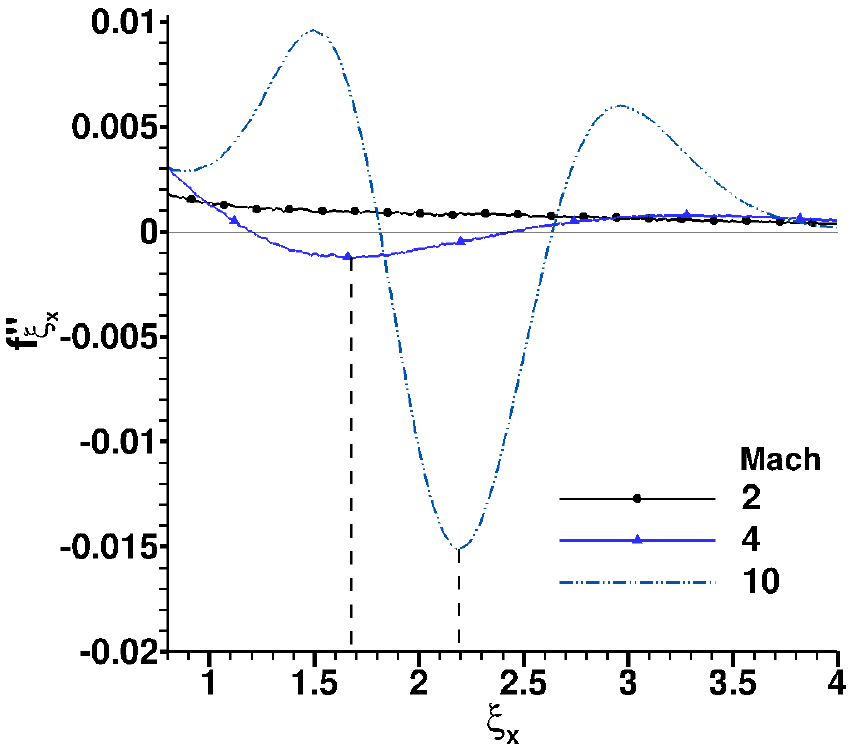}}}\,
    \sidesubfloat[]{\label{f:XixMversusf}{\includegraphics[width=0.44\textwidth]{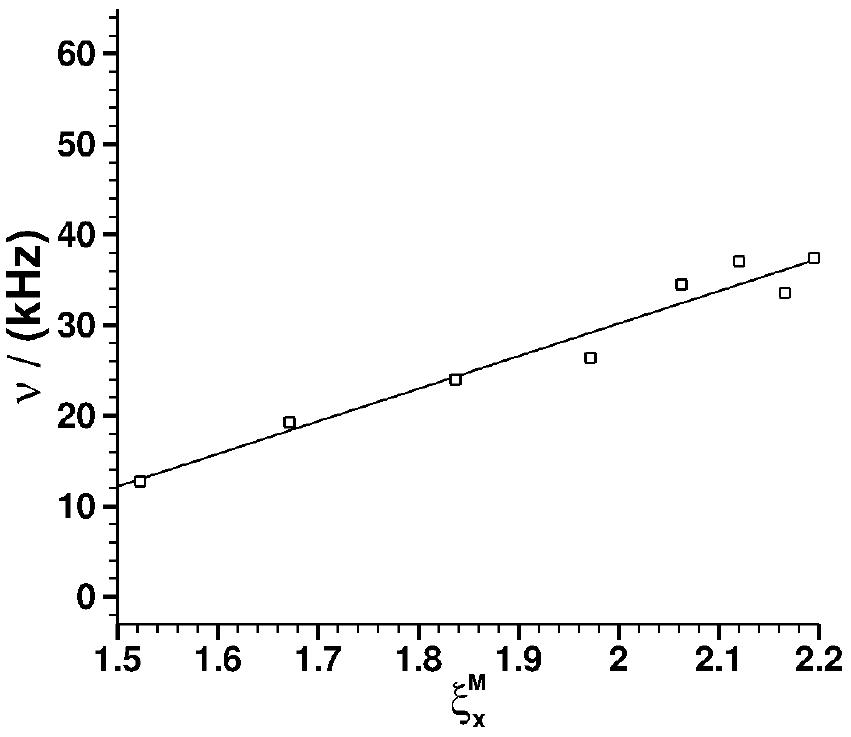}}}\,
    \sidesubfloat[]{\label{f:freqVersusVmp}{\includegraphics[width=0.44\textwidth]{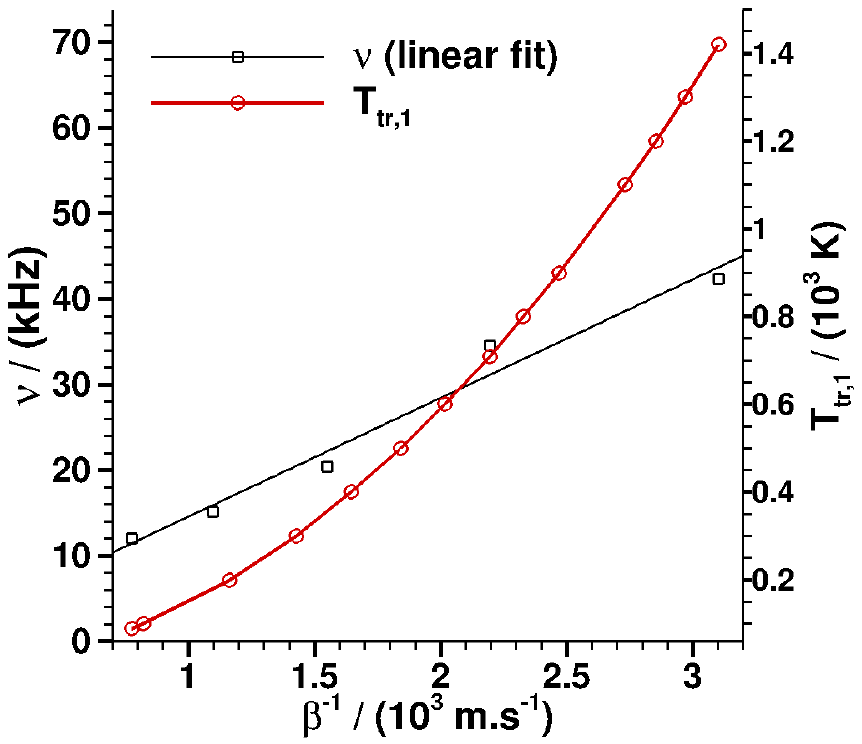}}}\,
    \caption{At probe $P$, (\textit{a}) analytical bimodal NCCS PDF, $f_{\xi_x}$, for Mach 2--10 and (\textit{b}) second derivative, $f''_{\xi_x}$, for Mach 2, 4, and 10. Location of $\xi_x^M$ in the energy space is indicated by verticle dashed-lines for Mach 4 and 10. $f''_{\xi_x}$ is smoothened using the Savitzky-Golay filter provided in the SciPy~\cite{SciPy} software.
 (\textit{c}) For Mach 3--10, a linear variation of DSMC-derived $\nu$ with $\xi_x^M$ of the PDFs shown in (\textit{a}). (\textit{d}) At Mach 7.2, for cases with varied input temperature, linear variation of DSMC-derived $\nu$ in the shock with $\beta^{-1}$ obtained from Mott-Smith bimodal velocity distribution. Also shown the variation of $\beta^{-1}$ with $T_{tr,1}$.}
    \label{f:UseOffxix}
\end{figure}

We begin with the correlation of the DSMC-derived average low-frequency of fluctuations in $\sigma_{xx}$ inside shocks with the change of shape of the analytical bimodal NCCS PDF of $f_{\xi_x}$ for Mach numbers ranging from three to 10 at $T_{tr,1}=710$~K.
Figure~\ref{f:bimodalfxix} shows the variation of the analytically-derived $f_{\xi_x}$ at the location of maximum bulk velocity gradient in the shock ($X/\lambda_1=0$) with Mach number.
At Mach 2, the shape of the PDF is very similar to a regular chi-squared distribution, indicating only a small contribution from high-energy upstream particles.
However, as the Mach number increases, their contribution increases leading to a prominant peak in the high-energy space, $\xi_x > 1$.
The DSMC-derived weighted average frequency is found to be linearly proportional to the location of this peak.
This is cosistent with the findings in our previous work~\cite{sawant2020kinetic} that the value of low-frequency depends on the collisions of low and high-energy particles from the subsonic downstream and supersonic/hypersonic upstream, respectively.
To determine the exact location of the peak, the second gradient of the PDF is evaluated, $f''_{\xi_x}$, as shown in figure~\ref{f:secGrad} for Mach 2, 4, and 10.
At Mach 4 and 10, two inflection points can be seen, as defined by $f''_{\xi_x}=0$, whereas at Mach 2, such an inflection point is absent.
Between the two inflection points, $f''_{\xi_x}$ attains a local minima, denoted as $\xi_x^M$, which corresponds to the respective peak location of $f_{\xi_x}$ in the high-energy space.
It is seen from figure~\ref{f:XixMversusf} that the DSMC-derived weighted-average low-frequency linearly changes with $\xi_x^M$. 
The data-points with increasing $\xi_x^M$ in figure~\ref{f:XixMversusf} correspond to Mach 3 to 10, which are also listed in Table~\ref{tab:XiXM}.
\vspace{\baselineskip}

\begin{figure}[H]
    \centering
    \sidesubfloat[]{\label{f:PSD}{\includegraphics[width=0.44\textwidth]{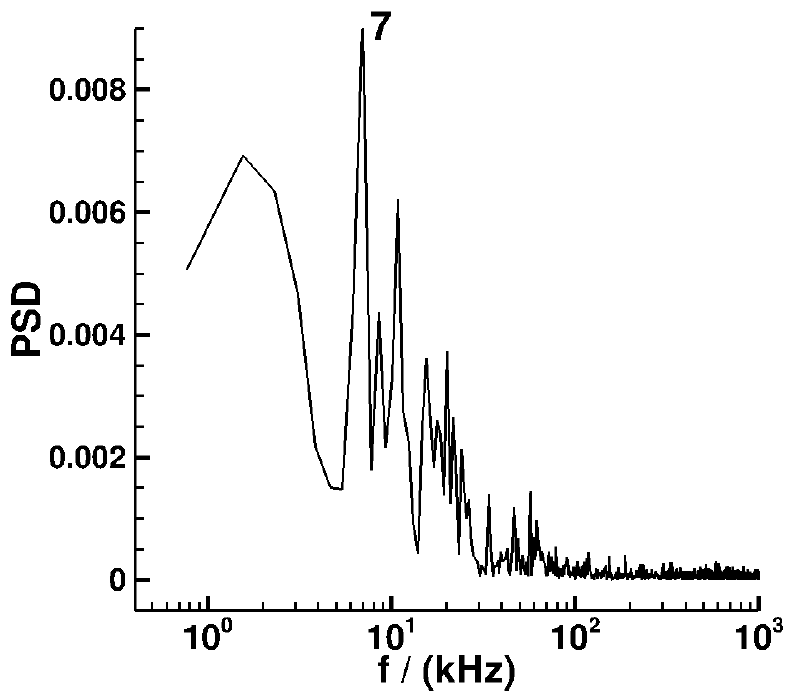}}}\hfill
    \sidesubfloat[]{\label{f:freqMatrix}{\includegraphics[width=0.44\textwidth]{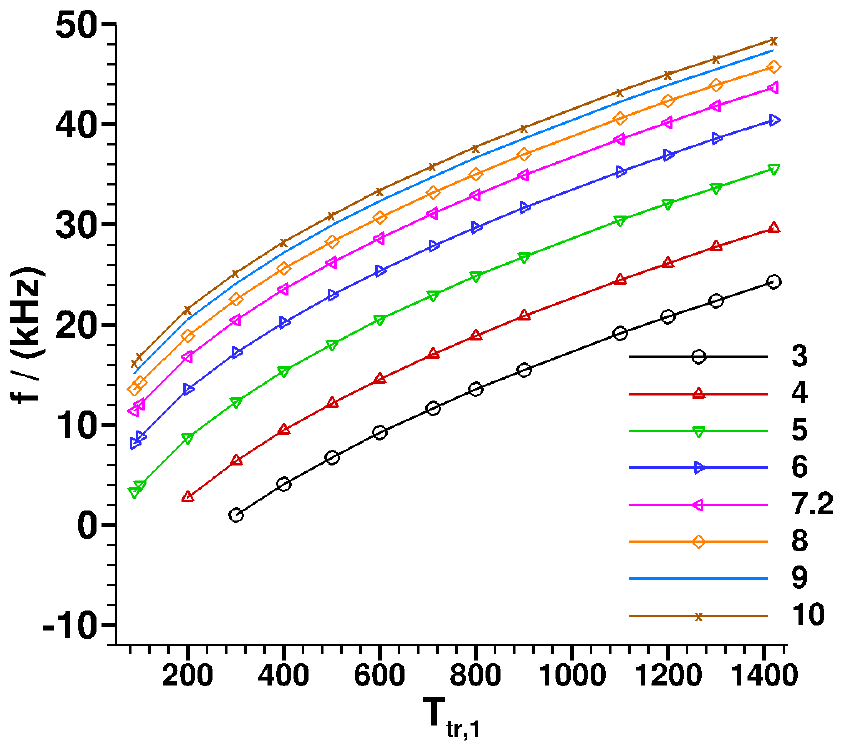}}}\,
    \caption{(\textit{a}) At M=4 and $T_{tr,1}=300$~K, the PSD obtained from the mean-subtracted data of $\sigma_{xx}\rho_1\beta_1^{-2}$ at the location of maximum bulk velocity gradient ($X/\lambda_1=0$). (\textit{b}) Predicted characteristic average low-frequencies as a function of input temperature for Mach numbers 3 to 10.}
    \label{f:prediction}
\end{figure}

Note, however, that the shape of $f_{\xi_x}$ does not change with changes in the upstream temperature, $T_{tr,1}$, at a given Mach number, because $\xi_x$ is normalized by $\beta^2$.
Therefore, the change in low-frequency due to the variation in freestream temperature cannot be estimated using only the linear fit in figure~\ref{f:XixMversusf}.
In this case, DSMC-derived $\nu$ is found to be linearly proportional to the most-probable speed, $\beta^{-1}$, inside a shock calculated from the analytical Mott-Smith velocity distribution function.
Figure~\ref{f:freqVersusVmp} shows such a linear variation at Mach 7.2 for data-points with increasing $\beta^{-1}$ corresponding to $T_{tr,1}/710$=1/8, 1/4, 1/2, 1, and 2. 
The variation of $\beta^{-1}$ obtained from Mott-Smith with $T_{tr,1}$ is also given in figure~\ref{f:freqVersusVmp}.
The aforementioned two approaches can be combined to determine $\nu$ between Mach 3 to 10 and $T_{tr,1}$ between 89 to 1420~K as, 
\begin{equation} 
\begin{split}
\nu(M^*,T_{tr,1}^*) &= \nu(M=7.2,T_{tr,1}^*) -\frac{d\nu}{d\xi_x^M}[\xi_x^M(M=7.2)-\xi_x^M(M^*)]\\
\end{split}
\label{eq:freq}
\end{equation}
where $M^*$ and $T_{tr,1}^*$ are the desired Mach number and freestream temperature, respectively, and the slope $d\nu/d\xi_x^M$ is obtained from figure~\ref{f:XixMversusf} to be equal to 36~kHz.
For example, using equation~\ref{eq:freq}, the predicted low-frequency at $M$=4 and $T_{tr,1}$=300~K is 6.4~kHz.
The frequency at $M$=7.2 and $T_{tr,1}^*=$300~K is 20.46~kHz, which is obtained by linear interpolation based on figure~\ref{f:freqVersusVmp} using the $Y$-intercept and slope of the fitted line equal to 0.6867~kHz and $d\nu/d\beta^{-1}=14.8~m^{-1}$, respectively.
In comparison, the peak-frequency in the normalized PSD of the DSMC-derived mean-subtracted instantaneous data of $\sigma_{xx}\rho_1\beta_1^{-2}$ is 7~kHz, as shown in figure~\ref{f:PSD}, in close agreement with our estimate.
As discussed in Ref.~\cite{sawant2020kinetic}, the PSD is obtained using the Welch's method~\cite{welch1967use,solomon1991psd} in the Scipy~\cite{SciPy} software with two Hann-window weighted segments of data sampled with a frequency of 100~MHz prior to the Fast Fourier Transform (FFT) such that the frequency resolution is 0.775~kHz.
Considering the complexity of this full simulation, equation~\ref{eq:freq} is a useful scaling relationship for extension to other Mach numbers and temperatures.
\begin{table}[H]
\begin{threeparttable}
\caption{$\xi_x^M$ for Mach numbers ranging from 3 to 10.}
\label{tab:XiXM}       
\begin{tabular}{|c|c|c|c|c|c|c|c|c|c|}
\hline\noalign{\smallskip}
\textbf{M} & 3 & 4 & 5 & 6 & 7.2 & 8 & 9 & 10     \\\hline
\textbf{$\xi_x^M$}   & 1.5224 & 1.6715 & 1.8366 & 1.9716 & 2.0622 & 2.1202 & 2.1654 & 2.195\\
\noalign{\smallskip}\hline\noalign{\smallskip}
\end{tabular}
\end{threeparttable}
\end{table}

Finally, using this approach, the average low-frequencies are predicted for input temperature ranging from 89~K to 1420~K for Mach numbers 3 to 10 and shown in figure~\ref{f:freqMatrix}.
It is seen that the chacteristic average low-frequency increases with increase in temperature and Mach number, however, the maximum range of frequencies is within 50~kHz for the parameter range examined.
This figure can be used to read out the average low-frequency fluctuation for a given input temperature and shock Mach number for researchers in hypersonic transition.

\vspace{-1em}
\section{Conclusion}~\label{sec:conclusion}
\vspace{-1em}

This work demonstrates a simple approach to estimate the low-frequency fluctuations in a 1-D shock structure of argon, using analytically derived PDFs of particle energies in local equilibrium. The latter compare well with DSMC-derived PDFs in the upstream and downstream equilibrium regions with respect to a one-dimensional shock.
It is shown for the first time that they have the form of NCCS PDFs.
Then, using the Mott-Smith model, the bimodal NCCS PDFs inside the nonequilibrium regions of shocks are constructed as a linear combination of the equilibrium upstream and downstream PDFs.
\footnote{The Appendix provides a python function to generate these PDFs.}
\vspace{\baselineskip}

The bimodal NCCS PDFs at the location in the shock where the velocity gradient is maximum ($X/\lambda_1=0$) are correlated with the low-frequencies of shock fluctuations previously obtained from DSMC for $3 \le M \le 10$.
It is found that the weighted-averaged low-frequency is directly proportional to the location of the peak in the PDF of $f_{\xi_x}$ generated by the contribution of particles from the upstream.
For cases where the Mach number is constant but input temperature is varied, the low-frequencies are found to be proportional to the most-probable-speed at $X/\lambda_1=0$ obtained from the Mott-Smith velocity PDF.
Using these linear functions, it is demonstrated that one can estimate the low-frequency of fluctuations for any arbitrary input condition not explicitly simulated in our previous work.
In the future, a similar approach can be used for other gases of practical importance, where the DSMC-derived low-frequencies can be obtained for a few selected input conditions and using those, a broader database of low-frequencies can be generated through linear correlations with analytical PDFs.
\vspace{\baselineskip}

The estimates provided for low-frequency unsteadiness generated at the shock can be used in receptivity and linear stabiliy analysis studies of laminar-turbulent transition of high-speed boundry layer flows, e.g. as boundary conditions in the analyses or to construct simplified models that account for the interaction of leading-edge shock with the boundary layer. The same estimates can also be used to aid understanding of the changes in the spectrum of freestream noise after it interacts with the shock prior to excitation of the boundary-layer.

\section{Acknowledgement}
The research conducted in this paper is supported by the Office of Naval Research under the grant No. N000141202195 titled, “Multi-scale modeling of unsteady shock-boundary layer hypersonic flow instabilities.”
This work used the STAMPEDE2 supercomputing resources provided by the Extreme Science and Engineering Discovery Environment (XSEDE) at the Texas Advanced Computing Center (TACC) through allocation TG-PHY160006.

\appendix
\section{Python Code for Generating the Bimodal NCCS PDFs}~\label{app:code}
This appendix gives the code snippet for generating bimodal NCCS distribution.
The values of code variables `R', `Ux1', `Ttr1', `beta1', `Ux2', `Ttr2', `beta2' are 208.243, 3572.24, 710, $\num{1.849e-3}$, 944.74, 12120.6, $\num{4.451e-4}$, respectively.
The values of code variables `Ux', `Ttr', `beta' are obtained from the Mott-Smith velocity distribution and equal to 2012.8, 10149.8, $\num{4.864e-3}$, respectively, at probe $P$.
The code variables `psi1' and `psi2' are Mott-Smith fractions and equal to 0.40649 and 0.59351, respectively, at probe $P$.
\begin{lstlisting}
#Code for generating a bimodal NCCS distribution
#Note:indices 0, 1, 2 of the following arrays
#is for the bimodal PDF of xi_x, xi_y, xi
#df1, df2: degrees of freedom
#nc1, nc2: noncentrality parameters

size=10000000
size1=int(psi1*size)
size2=int(psi2*size)

Var1 = R*Ttr1
s1=0.5*(Ttr1/Ttr) #scaling factor
beta1Sq=beta1*beta1
Ux1Sq=Ux1*Ux1
df1 = [1,1,3]
nc1 = [Ux1Sq/Var1, 0, Ux1Sq/Var1]

Var2 = R*Ttr2
s2=0.5*(Ttr2/Ttr)
beta2Sq=beta2*beta2
Ux2Sq=Ux2*Ux2
df2 = [1,1,3]
nc2 = [Ux2Sq/Var2, 0, Ux2Sq/Var2]

UxSq=Ux*Ux
betaSq=beta*beta
MeanSqBetaSq = [UxSq*betaSq, 0, UxSq*betaSq]

bins=3200
bin_edges_NCCS = np.zeros(bins+1)
bin_center_NCCS = np.zeros(bins)

i=0 #For the bimodal PDF of xi_x
mean, var, skew, kurt = ncx2.stats(df1[i], nc1[i],
                         scale=s1, moments='mvsk')
r1 = ncx2.rvs(df1[i], nc1[i], size=size1,scale=s1,
              loc=-MeanSqBetaSq[i])
mean, var, skew, kurt = ncx2.stats(df2[i], nc2[i],
                         scale=s2, moments='mvsk')
r2 = ncx2.rvs(df2[i], nc2[i], size=size2,scale=s2,
              loc=-MeanSqBetaSq[i])
r=np.concatenate((r1, r2))
hist_NCCS, bin_edges_NCCS = np.histogram(r,bins=bins,
                             density=True)
print(`mean, stddev', r.mean(), r.std())
fig, ax = plt.subplots(1, 1)
ax.hist(r,density=True,bins=bins,histtype=`stepfilled',
        alpha=0.4)
plt.show()
\end{lstlisting}

\newpage
\bibliographystyle{spphys}       
\bibliography{References}   


\end{document}